\documentclass[12pt,cite,a4paper]{article}
\usepackage {indentfirst}
\usepackage {graphicx}

\begin{document}
\baselineskip=18pt

\author{A.Loskutov, I.Istomin and O.Kotlyarov\\
{\small Physics Faculty, Moscow State University,
Moscow 119899 Russia}}

\title{DATA ANALYSIS: GENERALISATIONS OF THE LOCAL APPROXIMATION METHOD BY SINGULAR
SPECTRUM ANALYSIS}

\date{ }

\maketitle

\begin{abstract}

We study forecasting capabilities of the methods of
Singular Spectrum Analysis (SSA) and Local Approximation (LA). A practical
implementation of these methods to several time series is described. Details
of the algorithms of these methods are discussed. Advantages and
disadvantages of SSA and LA are given. On the basis of our results we
generalize LA and SSA and propose a new way for time series forecasting
including strongly noisy signals. This allowed us to extend the range of
application of LA in comparison with the known standard scheme. For the
problem of forecasting, the accuracy enhancement of the numerical
computations is discussed.

\end{abstract}

\vspace{0.3cm}

\section{Introduction}

Analysis of results of any observable is often based on the processing of
initial experimental data. In many cases these data represent a time series,
i.e. a certain sequence with elements in a chronological order. The
processing of the time series with the purpose of extraction of the useful
information about properties of the corresponding system is a very important
and interesting task. However, in many cases attention is paid not to
research of properties of the system, but to the forecast its further
dynamics. For example, in meteorology the practical interest refers first of
all to the weather forecast in a near future. Thus, besides the study of the
system properties, the task of the forecasting of the further trajectory is
also of practical significance. This problem does not exclusively belong to
meteorology, it also is very important both in geophysics and astrophysics.
It concerns the prediction of earthquakes, forecast of solar activity. In
financial analysis it concerns the forecast of share prices, exchange rates
etc.

The most widespread methods of the forecast of irregular (quasi periodic and
stochastic) time series are currently the methods of ARMA type (Auto
Regressive Moving Average) \cite{BoxJenk}, which have already for a long time been
applied in statistics, economy and meteorology \cite{MonPit}. The basic idea is
expression of the following elements in the time series by the previous
ones. This is probably a unique way, which can be used in the situation when
any other modeling of the system is not available, except for investigations
the observation data. The validity of such an approach was proven only after
the Takens paper at the beginning of the 80s \cite{Takens}.

Besides the validation of ARMA, the Takens ideas initiated the development
of new methods of analysis and forecasting the time series by the theory of
dynamical systems. One of such methods is a method of local approximation
(LA), offered for the first time in the work \cite{BFarSidfirst} for the forecast of the
chaotic time series. This method, which algorithm is briefly described in
the present work, has a number of advantages against a traditional method of
autoregression. However it is not yet widely spread basically because of
difficulties in application to short and noisy time series.

In the present paper we generalize the efficiency of LA for forecasting
noisy time series by means of a preliminary filtration of the data with the
help of Singular Spectrum Analysis (SSA) \cite{VautYiou}. The SSA method is also
developed within the framework of nonlinear dynamics, however it is used
mainly for definition of the basic components and suppression of noise \cite{GhilAl},
though there exist algorithms, based on it, which allow the forecast \cite{Caterpil}. We
hope that the offered generalization of the LA method should admit
considerably to expand the area of its application and to the forecast of
the chaotic and quasi-periodic behavior of nonlinear dynamical systems.

\bigskip

\section{Methods of processing of time series}

In this section we give some basis of the time-series analysis. This allows
us to generalize the LA method and propose a new way for forecasting chaotic
data.

\subsection{Method of delays}

A foundation of the majority of approaches related to the processing of time
series $\left\{ {x_{1} , \ldots ,x_{N}}  \right\}$ is the construction of a
set of delayed vectors (or vectors in the state space) $z_{i} = \left(
{x_{i} ,x_{i + 1} , \ldots ,x_{i + m}}  \right)^{T}$, $i = 1,\,2\,, \ldots
,N - m + 1$ \cite{BrooKing}. This is the first and necessary step in the methods of the
analysis of time series developed in the framework of nonlinear dynamics. In
a certain sense, the state space is equivalent to the phase space of the
corresponding nonlinear dynamical system, which generated the time series
[3, 9]. Thus, it has been proven that many-dimensional systems can be
described by time series, i.e. a scalar function of the system state
obtained in an experiment. In its turn, the possibility of the description
and reconstruction of the system dynamics allows, under the certain
conditions, to predict its further behavior \cite{LoskutovIK}.

As a rule, for the construction of vectors $z_{i} $ the method of delays is
used. This method similarly autoregression, establishes the transformation
from an initial one-dimensional (scalar) time series to a many-dimensional
(vector) representation. For this transformation, each many-dimensional
vector is formed from some number (say $m$) of successive values of the
initial time series. The result can be presented as a set of "photos" of a
series made through a window sliding along the series, in which only
\textit{m} consecutive elements may observe simultaneously:

\begin{equation}
\label{eq1}
X = {\begin{array}{*{20}c}
 {{\begin{array}{*{20}c}
 { \ \ \ \ \ \vdots}  \hfill \\
 {\ \ \ f_{m + 1}}  \hfill \\
\end{array}} } \hfill & {{\begin{array}{*{20}c}
 { \ \ \ \ \vdots}  \hfill \\
 {\ \ f_{m + 2}}  \hfill \\
\end{array}} } \hfill & {{\begin{array}{*{20}c}
 {} \hfill \\
 {} \hfill \\
\end{array}} } \hfill & {{\begin{array}{*{20}c}
 {} \hfill \\
 {} \hfill \\
\end{array}} } \hfill \\
 {\left( {\left[ {{\begin{array}{*{20}c}
 {f_{m}}  \hfill \\
 { \ \vdots}  \hfill \\
 {f_{2}}  \hfill \\
 {f_{1}}  \hfill \\
\end{array}} } \right]} \right.} \hfill & {\left[ {{\begin{array}{*{20}c}
 {f_{m + 1}}  \hfill \\
 { \ \ \vdots}  \hfill \\
 {f_{3}}  \hfill \\
 {f_{2}}  \hfill \\
\end{array}} } \right]} \hfill & { \cdots}  \hfill & {\left. {\left[
{{\begin{array}{*{20}c}
 {f_{N}}  \hfill \\
 { \ \ \ \vdots}  \hfill \\
 {f_{N - m + 2}}  \hfill \\
 {f_{N - m + 1}}  \hfill \\
\end{array}} } \right]} \right)} \hfill \\
 {{\begin{array}{*{20}c}
 { \ \ \ \ \ \downarrow}  \hfill \\
 {} \hfill \\
\end{array}} } \hfill & {{\begin{array}{*{20}c}
 { \ \ f_{1}}  \hfill \\
 { \ \ \ \downarrow}  \hfill \\
\end{array}} } \hfill & {} \hfill & {{\begin{array}{*{20}c}
 {\ \ f_{N - m}}  \hfill \\
 { \ \ \ \ \ \vdots}  \hfill \\
\end{array}} } \hfill \\
\end{array}} .
\end{equation}

Here $f_{1} ,f_{2} , \ldots ,\,f_{N} $ are values of elements in the series
at time $t = 1,2, \ldots ,N$. Each square bracket is a vector in
$m$-dimensional space of delays; the sequence of such vectors gives an
observation matrix $X_{m \times \left( {N - m + 1} \right)} $, where
\textit{N} is the number of elements of an initial series. This matrix, in
every column of which there are parts of the same series moved relative to
each other, is many-dimensional form of an initial scalar series in space of
delays.

In a discrete case the described many-dimensional representation is given by
one parameter. This parameter is the dimension of the space of delays, or
the embedding dimension $m$. As shown, in many cases the opportunity of the
exact and reliable forecast depends on its correct choice. From the theory
of dynamical system it follows that $m \ge 2d + 1$, where $d$ is the
attractor dimension of the system, which generated the series. However this
condition is not constructive in a choice of the embedding. The most popular
algorithm for an estimation of the embedding dimension (and the system
dimension) is a Grassberger-Procaccia algorithm \cite{GrasProc}, but it is also not so
efficient for short (up to $10^{4}$ elements) time series.

For periodic and quasi-periodic time series there is an empirical rule
according to which it is necessary to choose $m \in \left( {{{T}
\mathord{\left/ {\vphantom {{T} {5}}} \right. \kern-\nulldelimiterspace}
{5}},\;T} \right)$, where $T$ is an average period \cite{GhilAl}.

The specified complexities in the determination of the embedding dimension
can be overcome by means of its definition within the framework of a used
forecast method. In this case an available series is divided into two
unequal parts. One of them (the smaller) is used for the quality check of
the forecast made on the basis of the other. The dimension, where the
forecast turns out to be the best, is considered optimal for the given
series.

\subsection{Method of local approximation (LA)}

Today the methods of autoregression (in the class of ARMA methods \cite{BoxJenk}) are
most frequently used for forecasting the time series. The autoregression
model of the order $p$-AR ($p$) has the following form:

\begin{equation}
\label{eq2}
f_{t} = a_{0} + a_{1} f_{t - 1} + a_{2} f_{t - 2} + \ldots + a_{p} f_{t - p}
.
\end{equation}

In this case, to predict the further trajectory of a series, it is necessary
first to determine the order of autoregression, and then on the basis of the
available data to obtain estimations of autoregression coefficients $\left\{
{a_{0} , \ldots ,a_{p}}  \right\}$. Here, however there is no unique
algorithm of choice of the order, and as a rule, it is chosen due to the
type of autocorrelation function. The corresponding coefficients are
estimated for all available data and assumed to be constant for any $t$.
Thus, the given approximation is a global one.

In general, the methods of global approximation give a quite good
description of the function when the number of free parameters are enough.
However, if the function is rather complex, there is no guarantee that we
can find such representation which allows to approximate efficiently the
analyzed function. In this way we can obtain an exclusive circle: The more
complex function, the more parameters are necessary, the more parameters
should be estimated, the more data are required. Increase of the number of
the used data (which is not always accessible) for a quite complex function
requires introduction of additional parameters to the model.

To avoid this cycling one can use LA. Its basic idea is to divide the domain
of the function into some local subdomains, construct approximate models and
estimate parameters separately in each area. If the function is smooth then
the subdomains can be small enough such that the function in each of them
does not change too sharply. It allows to apply in each domain simple models
(say, a linear one). The main condition of the efficient application of LA
is the correct choice of the size of a local area or, which is practically
the same, the number of the neighbors.

The method of LA \cite{BFarSidfirst} historically has become the first local method
developed for the forecast of time series on the basis of the Takens theory.
This method also applies representation (\ref{eq2}), but in this case it has three
basic differences from a method of autoregression:

\begin{itemize}
\item
coefficients in expression (\ref{eq2}) are estimated separately for each
of local areas;
\item
expression (\ref{eq2}) can include also nonlinear members, i.e. various
degrees of values of the series in the previous time (that is said to be an
approximation order);
\item
embedding dimension is used as a value of the autoregression
order.
\end{itemize}

Thus, the LA method turns out to be more valid for the choice "of the
autoregression order" and more flexible in use of the initial data.

Let us stop briefly on the basic steps of algorithm of the LA method.

{\bf 1. The choice of local representation}

This step includes evaluation of the embedding dimension and construction of
many-dimensional representation of a series, i.e. the matrix of delays $X$.

{\bf 2. Determination of a vicinity, i.e. the number of neighbors}

As a rule, the vicinity is given by a choice of number of the neighbors
$N^{s}$. For this purpose in the state space (among the columns of the
matrix $X$) the most "close" states are chosen. As the most simple criterion
of the closeness it is possible to choose the following: for the given
metric $\left\| {} \right\|$ and for \textit{the given} number of neighbors
$N^{s}$ the set $\left\{ {y^{t}} \right\}$ will be a set of the nearest
neighbors $x$, if $\sum\limits_{t = 1}^{N^{s}} {\left\| {y^{t} - x}
\right\|} \to min$. It should be noted that the closeness of $y^{t}$ to $x$,
though we consider the change dynamics in the observable, does not mean the
closeness in time.

{\bf 3. Choice of an approximation model and its identification}

In this step, for the determined embedding dimension the order of
approximation is chosen (usually either linear approximation or square-law
approximation in the previous instants are used):

\begin{equation}
\label{eq3}
x_{1}^{t + T} = f^{t + T}\left( {x^{t}} \right) = {\rm P}_{q} \left(
{x_{1}^{t} ,x_{2}^{t} , \ldots ,x_{m}^{t}}  \right),
\end{equation}

\noindent
where $T = 1,2, \ldots $, $q$ are the polynomial degree, i.e. the
approximation order. Therefore schemes of LA can be classified according to
the approximating polynomial order. Coefficients of the model are estimated
by a Singular Value Decomposition (SVD) \cite{Kah}.

In cases of strongly limited length of a series the approximation of a zero
order is used. Then the "forecast" depends only on the hit in the concrete
local area. Such a situation can be illustrated by the weather forecast
\cite{FarSidPred}.

{\bf 4. Forecasting}

The last step is the forecast for the next elements of the series. This is
made on the assumption that the evolution of the last elements occurs
according to the same law, as for other vectors from their local vicinity.

To make a forecast for a few steps forward, two basic ways of extrapolation
\cite{FarSidPred} are used. In the first or \textit{iterative} way for $T$ = 1 the model
parameters in (\ref{eq3}) are estimated, and the further forecast, $T$ = 2, 3, …,
represents a sequence of iterations. This means that the predicted value is
added to an initial series. Then, assuming that the obtained new vector in
the state space are in the same vicinity (that means that it evolves by the
same law), the forecast for one step forward is made. And so on.

The second, an alternative way, is \textit{direct} forecast. It consists of
the estimating of parameters separately for each $T$. This method allows to
make more exact forecast, since in this case there is no accumulation of an
error on each step \cite{LoskutovIKprep}.

By development of a LA method it was supposed, that the accuracy of
predictions is limited by the quality of approximation, which for an
available data set is determined by a number of points. However in many
cases the accuracy of a forecast can be limited as well by noise: even if we
know precisely the equations, the noise narrows limits of the
predictability. It brings an error in the determination of the initial
conditions and smears trajectories. In \cite{FarSidPred} it is shown that the influence
of noise on the quality of the forecast is very similar (by the
consequences) to the errors in approximation. In LA the fatal error arising
under influence of noise appears as a choice of neighbors. Thus, it is
necessary to generalize LA and overcome this difficulty.

\subsection{Singular spectral analysis}

Firstly the SSA method was developed for extraction periodic and
quasi-periodic components from time series. It was shown that this method
can be used for the improvement of a signal-to-noise ratio \cite{GhilAl}. Moreover,
recently there appeared options for the extension of SSA resources allowing
to make on this basis the forecast of further dynamics of a series. However,
in the present paper this method is used only for suppression of noise.

The basic idea of a SSA method consists of the transformation of the matrix
$X$ by the algorithm close to a method of principal components (PC). The use
of PC is the most important part of SSA which distinguish it from other
methods of nonlinear dynamics applied for the analysis and the forecast of
time series.

The main point of PC is a decrease of the dimension of initial space of the
factors (in this paper this is the space of delays) by means of the passage
to more "informative" variables (coordinates). As a result, new variables
are called the principal components (PCs). This passage is carried out via
orthogonal linear transformation.

In practice, in order to pass to PC it is necessary to calculate eigenvalues
and eigenvectors of the matrix $XX^{T}$. The last ones are chosen as a new
basis:

\begin{equation}
\label{eq4}
\Lambda = V\left( {XX^{T}} \right)V^{T},
\end{equation}

\noindent
where $\Lambda = \left( {{\begin{array}{*{20}c}
 {\lambda _{1}}  \hfill & {} \hfill & {} \hfill & {0} \hfill \\
 {} \hfill & {\lambda _{2}}  \hfill & {} \hfill & {} \hfill \\
 {} \hfill & {} \hfill & { \ddots}  \hfill & {} \hfill \\
 {0} \hfill & {} \hfill & {} \hfill & {\lambda _{M}}  \hfill \\
\end{array}} } \right)$ is a matrix of eigenvalues, $V_{M \times M} $ is a
matrix of eigenvectors. In addition, the matrix of PCs is $Z = VX$. At such
a transformation, PCs represent certain sets of point projections of an
initial set into eigenvectors. The eigenvalues characterize the scatter of
points along new axes. Usually they are ordered in the descending order.

The PCs have many useful properties. In SSA the resulting decomposition is
used for the extraction of the most significant components and the
truncation of random perturbations in the investigated series. The basic
idea of such a filtration is the use for reconstruction not all PCs of a
matrix $X$ but only the most significant ones. The significance of PCs is
usually determined by the values of their eigenvalues.

In general, approximation of a matrix $X$ imply the following:

\begin{equation}
\label{eq5}
\hat {X} = V_{r \times r}^{T} V_{r \times r} X,
\end{equation}

\noindent
where $V_{r \times r} $ is a part of eigenvector matrix corresponding to $r$
first principal components. After approximation of the matrix $X$ it is
necessary to reconstruct an initial time series. Reconstruction of the
matrix $X$ by the first principal components leads to the lost of the
initial diagonal image. That is the reason why at the reconstruction of the
initial time series it is necessary to average the matrix elements along
diagonals with the originally identical values (sf. (\ref{eq1})):

\begin{equation}
\label{eq6}
\hat {f}_{t} = \left\{ {{\begin{array}{l}
 {\frac{{1}}{{t}}\sum\limits_{i = 1}^{t} {\hat {x}_{M - t + i,t}}  ,\quad
\quad \quad \quad \;t < M,} \\
 {\frac{{1}}{{M}}\sum\limits_{i = 1}^{M} {\hat {x}_{M - t + i,t}}  ,\quad
\quad \quad \;\;M \le t \le N - M + 1,} \\
 {\frac{{1}}{{N - t + 1}}\sum\limits_{i = 1}^{N - t + 1} {\hat {x}_{i,t - M
+ i}}  ,\,\,N - M + 1 < t \le N,} \\
\end{array}} } \right.
\end{equation}

Thus, the algorithm of reconstruction of time series by SSA includes three
basic steps:
\begin{enumerate}
\item
Construction of the matrix $X$.
\item
Calculation of principal components and choice from them the most
significant.
\item
Reconstruction of the time series by the chosen principal components.
\end{enumerate}

The application of this algorithm allows to smooth an initial series,
decrease the noise level and raise a signal-to-noise ratio. However, the
methods of forecasting \cite{Caterpil} developed on its basis are insufficiently
effective for unperiodical time series. Like the LA method, SSA has some
modifications related to a preliminary centering and/or normalization of
rows in the matrix $X$. We shall use the variant with the centered rows
because it is more close to the PC standard.

\section{Generalization of LA by SSA. SSA-LA method.}

In this section we present a certain generalization of the described methods
via unification of some possibilities of LA and SSA.

\subsection{Algorithm of SSA-LA}

In the case of highly noisy series increasing the number of observation does
not allow to apply LA algorithm effectively because it is highly probably
that false neighbors will appear and, simultaneously true neighbors will be
eliminated. Therefore, to improve the quality of the forecast it is
necessary to combine both analyzed methods (SSA and LA). In this case SSA
will be used only for the filtration of an initial time series (i.e. noise
reduction), and the forecast will be made according to the LA method. Thus,
the proposed method (SSA-LA) can be considered as a generalization of LA
which allows to process highly noisy time series.

The SSA-LA algorithm consists of two basic stages:

\begin{enumerate}
\item
SSA-filtration for noise reduction;
\item
Forecasting of the modified time series by the LA method.
\end{enumerate}

It should be noted that to make a forecast a LA variant of the first order
and SSA with a centering were chosen. To estimate the coefficients in (\ref{eq2})
the PC method was used instead of SVD (see \cite{LoskutovIKprep}). This allows to raise the
accuracy of the forecasts.

Application of SSA and LA shows that at the SSA reconstruction the first and
the last $M$ elements of the series can be found with quite large deviations
from initial values, whereas other parts of the series are approximated much
more precisely. Apparently, decreasing the accuracy of boundary value
approximations is a result of a shorter interval of the averaging used in
(\ref{eq6}). Therefore for more qualitative forecasting the first and the last $M$
values were truncated, i.e. the forecast was made starting from $N - M + 1$
value.

\subsection{Numerical results}

In this section some numerical results are presented. They can help to
estimate the opportunities of SSA-LA application for the forecast of highly
noisy irregular time series. These time series were generated on the basis
of the known Mackey-Glass equation with a delay (\ref{eq7}) and $x$-components of
finite-difference approximation of the Lorenz system (\ref{eq8}). Uncorrelated Gauss
noise with zero expectation were added.

\begin{equation}
\label{eq7}
x_{i + 1} = \left( {1 - b} \right) \cdot x_{i} + \frac{{a \cdot x_{i -
\Delta T}} }{{1 + \left( {x_{i - \Delta T}}  \right)^{c}}},
\quad
a = 0.2,\;\;b = 0.1,\;\;c = 10,\;\;\Delta T = 17,
\end{equation}

\begin{equation}
\label{eq8}
\left\{ {{\begin{array}{l}
 {\dot {x} = \sigma \left( {y - x} \right)} \\
 {\dot {y} = \left( {r - z} \right)x - y} \\
 {\dot {z} = xy - bz} \\
\end{array}} } \right.,
\quad
\sigma = 5,\;\;r = 15,\;\;b = 1,\;\;\Delta t = 0.02.
\end{equation}

It should be noted that base time series were designed once for all
numerical analysis, and a noise component was generated independently for
each experiment.

\begin{figure}[htbp]
\centering
\includegraphics*[bbllx=0.19in,bblly=0.18in,bburx=5.59in,bbury=3.85in,scale=0.69]{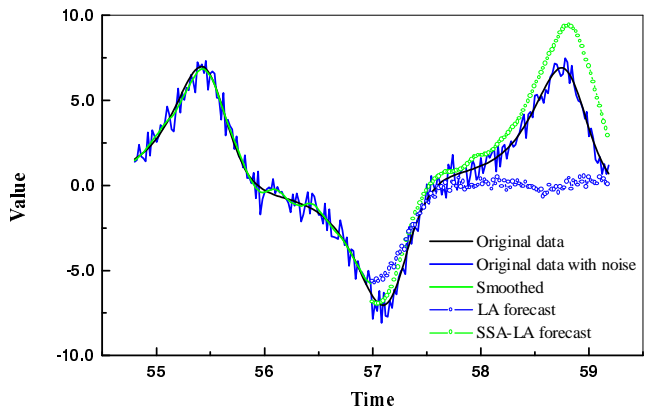}
\caption{ \small SSA-filtering and forecast for $x$-component for the Lorenz
system with 5\% noise. The forecast start point is 57.0.}\label{LorPred}
\end{figure}

\begin{figure}[htbp]
\centering
\includegraphics*[bbllx=0.19in,bblly=0.18in,bburx=5.41in,bbury=3.75in,scale=0.69]{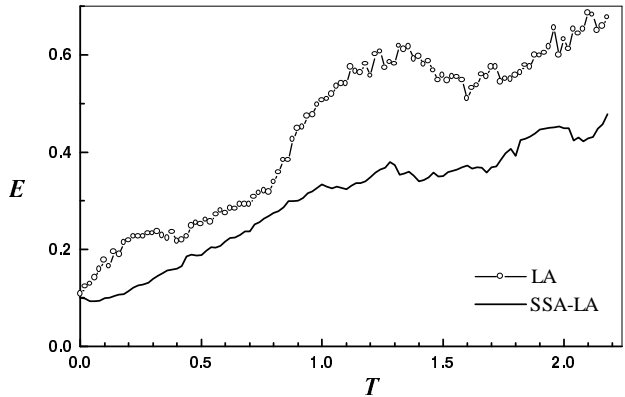}
\caption{\small Prediction error for $x$-component of the Lorenz system
with 5\% noise. Averaging over 500 forecasts.}
\label{Lorenz}
\end{figure}

\begin{figure}[htbp]
\centering
\includegraphics*[bbllx=0.19in,bblly=0.18in,bburx=5.41in,bbury=3.75in,scale=0.69]{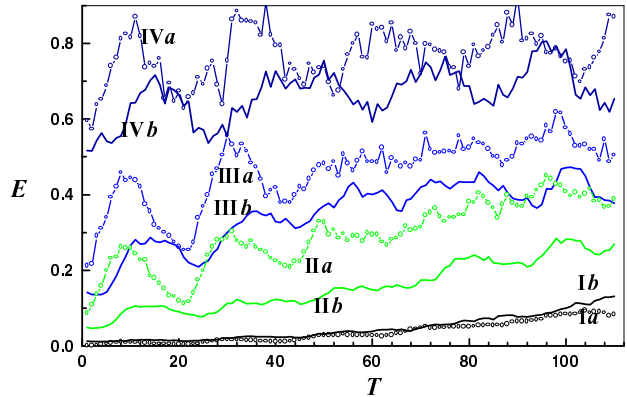}
\caption {\small The prediction error for the Mackey-Glass
equation as a function of the forecast length without noise (I) and with
noise of the amplitude of 1.5\% (II), 5\% (III), 20\%(IV). LA (\textit{a})
and SSA-LA (\textit{b}). Average over 500 forecasts.}
\label{McGFull}
\end{figure}

The results of numerical analysis are given in Figs.~\ref{LorPred},\ref{Lorenz},\ref{McGFull}. For the forecast,
3600 points of each time series were chosen, and 110 points from them were
used only for an estimation of the forecast quality. Parameters of a
SSA-filtration are the following: $M = 18$, $r = 6$. LA parameters are: $m =
6$, $N^{s} = 21$.

First let us consider the example of $x$-component forecast for the Lorenz
system (Fig.~\ref{LorPred}). The left part of the diagram (up to the point 57.0) shows
the result of a SSA-filtration of a noisy signal (last 110 points), which is
a quite accurate reconstruction of the initial series (without noise).
However, the phase is reconstructed with a small enough error. From the
point 57.0 the forecast begins (circles). One can see that the preliminary
filtration allows to raise essentially the accuracy of the forecast. However
basically the accuracy of the forecast can depend on the moment of the
beginning. To get more real estimation of the forecast quality by LA and
SSA-LA it is necessary to apply the characteristic which is not depends on
the start point. Following \cite{FarSidPred} such a characteristic is the value of the
standard root-mean square error or the prediction error:

\begin{equation}
\label{eq9}
E = \sqrt {\frac{{\left\langle {\left( {f_{t + T} - \hat {f}_{t + T}}
\right)^{2}} \right\rangle _{t}} }{{\left\langle {\left( {f_{t} -
\left\langle {f_{t}}  \right\rangle _{t}}  \right)^{2}} \right\rangle _{t}
}}} .
\end{equation}

\bigskip

Here the averaging is made by the moments of the beginning of the forecast.
If $E$ exceeds one then the forecast is not successful, and would be better
to exploit the average value as the forecast. For more reliability of the
error estimation, expression (\ref{eq9}) contains median instead of the arithmetic
mean for the initial moments of the forecast.

For the same time series, Fig.~\ref{Lorenz} shows the prediction error as a function of
the forecast length. Like the example in Fig.~\ref{LorPred}, the mean forecast with a
preliminary filtration is better. The error in the SSA-LA case is, as a
rule, less than the error for the standard LA. Dependencies shown in Fig.~\ref{Lorenz}
were constructed for various amplitudes of noise. Except for zero noise
level, the forecast was more exact for the LA-SSA method.

By the same criterion, the series obtained from the Mackey-Glass equation
have been analyzed. The results turn out to be very similar to the Lorenz
series Fig.~\ref{McGFull}. The observed advantage of the LA method in the absence of
noise is a natural result of quite small distortions in the initial series
obtained during the SSA-filtration. However for the noisy series this fact
is completely compensated by suppression of noise. So, the forecast by the
SSA-LA method is more precise than the standard LA-approach.

Thus, the numerical analysis show that the preliminary SSA-filtration allows
us to raise considerably the accuracy of the forecast obtained by the LA
method. In turn, combination of LA-SSA can be very useful for the study and
forecast the noisy time series when the standard LA is not efficient.

\bigskip
\section{Conclusion}

The forecast of time series by the LA method is currently more an
opportunity than a real researching tool. At the same time, the LA algorithm
has a certain advantage over the usual autoregression. This advantage
consists of the use of piecewise-linear approximation instead of a
global-linear one. This allows to predict irregular time series for which a
linear autoregressive representation can not be applied. The basic reason
which limits the use of the LA method is that its efficient application is
possible only for the forecast of a sufficiently long time series. In the
case of a high noise level the requirements to the length of a series
essentially grow; in quite rare situations one can find a series of the
necessary length.

To decrease this limitation in the present paper we propose to use a
preliminary filtration of a series by the SSA method. As known, SSA is a
good tool for noise suppression especially for irregular time series, when
the Fourier-filtration cannot be applied. It is numerically shown
combination of SSA and LA gives more accurate forecast than in the standard
LA method. This result practically did not depend on the noise level (except
for the case of its absolute absence), on the length of the forecast and on
the nature of the system which generated the analyzed series. It seems to be
true that some advantage of LA over SSA-LA in the absence of noise is
completely compensated by the significant advantage of the SSA-LA for the
noisy data because as a rule, real data includes a certain noise. Thus, one
can say that the use of the SSA-filtration as a necessary component can
essentially extend the area of the LA application.

\end{document}